\documentclass[a4paper]{jpconf}

\usepackage{amssymb}
\usepackage{amsmath}

\usepackage{latexsym}
\usepackage{graphicx}

\usepackage{cite}

\newcommand{\be}{\begin{equation}}
\newcommand{\ee}{\end{equation}}

\newcommand{\dlt}{\delta}

\newcommand{\bt}{\beta}

\newcommand{\al}{\alpha}
\newcommand{\ra}{\rightarrow}
\newcommand{\sgm}{\sigma}

\newcommand{\Gm}{\Gamma}

\newcommand{\lbd}{\lambda}

\newcommand{\rgl}{\rangle}
\newcommand{\lgl}{\langle}

\begin{document}

\title{Quantum theory of measurements as quantum decision theory}

\author{V I Yukalov$^{1,2}$ and D Sornette$^{2,3}$}

\address{$^1$Bogolubov Laboratory of Theoretical Physics, 
Joint Institute for Nuclear Research, \\ Dubna 141980, Russia \\
$^2$DMTEC, ETH Z\"urich, Swiss Federal Institute of Technology, 
Z\"urich CH-8092, Switzerland \\
$^3$Swiss Finance Institute, c/o University of Geneva, 
Geneva CH-1211, Switzerland}

\ead{yukalov@theor.jinr.ru}

\begin{abstract}
Theory of quantum measurements is often classified as decision theory. An
event in decision theory corresponds to the measurement of an observable.
This analogy looks clear for operationally testable simple events. However,
the situation is essentially more complicated in the case of composite events.
The most difficult point is the relation between decisions under uncertainty
and measurements under uncertainty. We suggest a unified language for
describing the processes of quantum decision making and quantum measurements.
The notion of quantum measurements under uncertainty is introduced. We show
that the correct mathematical foundation for the theory of measurements under
uncertainty, as well as for quantum decision theory dealing with uncertain
events, requires the use of positive operator-valued measure that is a
generalization of projection-valued measure. The latter is appropriate for
operationally testable events, while the former is necessary for characterizing
operationally uncertain events. In both decision making and quantum
measurements, one has to distinguish composite non-entangled events from
composite entangled events. Quantum probability can be essentially different
from classical probability only for entangled events. The necessary condition
for the appearance of an interference term in the quantum probability
is the occurrence of entangled prospects and the existence of an entangled
strategic state of a decision maker or of an entangled statistical state of
a measuring device.
\end{abstract}

\section{Introduction}

Developing the theory of quantum measurements, von Neumann \cite{Neumann_1}
mentioned that the process of quantum measurements is analogous to decision
making. The formal analogy between these processes has been described in
several mathematical works \cite{Benioff_2,Holevo_3,Holevo_4,Holevo_5}.
However, this analogy remained rather formal, without comparing quantum
measurements with real decision making, as done by humans. Several important
questions have not been answered:

(i) First of all, in what sense human decision making could be characterized
by quantum measurements?

(ii) What would be a general scheme for describing measurements under
uncertainty and decisions under uncertainty?

(iii) How to correctly define a quantum probability of non-commuting events
for both, quantum measurements and quantum decisions?

(iv) What is the role of entanglement in quantum measurements, and in quantum
decision making?

(v) When should decision making be treated by quantum rules and when is it
sufficient to use classical theory?

In this report, we present a general approach and mathematical techniques
common to quantum measurements and decision making that provides natural
answers to these questions. The reason for developing a common quantum
approach to measurements and to decision making is twofold. First, quantum
theory provides tools for taking into account behavioral biases in human
decision making \cite{Yukalov_6,Yukalov_7,Yukalov_8}. Note that the possibility
of describing cognition effects by means of quantum theory was suggested by Bohr
\cite{Bohr_9,Bohr_10}. Second, formulating a quantum theory of decision making
defines the main directions for creating artificial quantum intelligence
\cite{Yukalov_11,Yukalov_12}.

\section{Operationally testable events}

Let us first briefly recall the definition of quantum probabilities for
measurements corresponding to operationally testable events and their
connection to simple events in decision making
\cite{Yukalov_6,Yukalov_7,Yukalov_8}.

The operator of an observable $\hat{A}$ is a self-adjoint operator, whose
eigenvectors are given by the eigenproblem
\be
\label{1}
 \hat A | n \rgl = A_n | n \rgl \;  ,
\ee
forming a complete basis. The closed linear envelope of this basis composes
a Hilbert space $\mathcal{H}_A \equiv span\{\vert n \rangle\}$. The measurement
of an eigenvalue $A_n$ is an event that we denote by the same letter. This
event is represented by a projector $\hat{P}_n$ according to the correspondence
\be
\label{2}
 A_n \ra \hat P_n \equiv | n \rgl \lgl n | \;  .
\ee
The system statistical state, or the decision maker strategic state, is given
by a statistical operator $\hat{\rho}_A$. The probability of an event $A_n$ is
\be
\label{3}
 p(A_n) \equiv {\rm Tr}_A \hat \rho_A \hat P_n \equiv \lgl \hat P_n \rgl \;  .
\ee
In the theory of measurements or decision theory, the set
$\mathcal{A} \equiv \{\hat{P}_n\}$ of projectors plays the role of the algebra
of observables, with the expected value (3) being the event probability.

In eigenproblem (1), a nondegenerate spectrum is tacitly assumed. If the
spectrum is degenerate, then in the eigenproblem
\be
\label{4}
\hat A | n_j \rgl = A_n | n_j \rgl
\ee
an eigenvalue $A_n$ corresponds to several eigenvectors $\vert n_j \rangle$,
where $j = 1,2,\ldots$. The Hilbert space can again be composed spanning all
these eigenvectors, $\mathcal{H}_A \equiv \{\vert n_j \rangle\}$. Now an event
of measuring $A_n$ is associated with the projector
\be
\label{5}
 \hat P_n \equiv \sum_j \hat P_{n_j} \qquad
( \hat P_{n_j} \equiv  | n_j \rgl \lgl n_j | ) \; .
\ee
And the event probability is defined as in Eq. (3).

A basis, formed by the eigenvectors of a self-adjoint operator, is complete,
because a self-adjoint operator is normal. The set of eigenvectors of any
normal operator, such that $\hat{A}^+ \hat{A} = \hat{A} \hat{A}^+$, forms a
complete basis. Normal operators include self-adjoint and unitary operators.
In the case of degenerate spectra, the basis may be not uniquely defined. In
quantum measurements, this does not lead to any principal problem, since one
can use the summary projector (5). But in decision theory, the question arises:
in what sense a single event $A_n$ can correspond to several projectors
$\hat{P}_{n_j}$, since an operationally testable event either happens or not?

Fortunately, this problem of nonuniqueness is easily avoidable, both in the
theory of quantum measurements and in decision theory. This is done by invoking
the von Neumann suggestion of degeneracy lifting \cite{Neumann_1}. For this
purpose, the operator of the observable $\hat{A}$ is slightly shifted by an
infinitesimal term,
\be
\label{6}
 \hat A \ra \hat A + \nu \hat\Gm \qquad (\nu \ra 0 ) \;  ,
\ee
where $\hat{\Gamma}$ is an operator lifting the symmetry responsible for the
spectrum degeneracy. Then the operator spectrum splits into the set
\be
\label{7}
 A_n \ra A_n + \nu \Gm_{n_j} \;   ,
\ee
thus, removing the degeneracy. Finally, the event probability is defined as
\be
\label{8}
  p(A_n) = \lim_{\nu\ra 0} p(A_n + \nu \Gm_{n_j} ) \; .
\ee
In that way, the degeneracy is avoided, and one always deals with a unique
correspondence between eigenvalues, representing events, and eigenvectors.
This procedure is also analogous to the Bogolubov method of quasi-averages,
breaking the symmetry of statistical systems by introducing infinitesimal
sources \cite{Bogolubov_13,Bogolubov_14}.

The union of mutually orthogonal (incompatible) events is represented by
the projector sum:
\be
\label{9}
 \bigcup_n A_n \ra \sum_n \hat P_n \qquad
( \hat P_m \hat P_n = \dlt_{mn} \hat P_n ) \;  .
\ee
The probability of such a union is additive:
\be
\label{10}
 p \left ( \bigcup_n A_n \right ) = \sum_n p( A_n) \;  .
\ee

In the definition of the quantum probability (3), the averaging is done
with a statistical operator $\hat{\rho}_A$, generally, implying a mixed
state. In some special cases, a quantum system can be prepared in a pure
state described by a wave function $\vert \psi \rangle$, which corresponds
to the statistical operator $\vert \psi \rangle \langle \psi \vert$. The
setups of experiments with physical systems and with decision makers are
quite different. Accomplishing quantum measurements, we may meet two types
of systems, open and quasi-isolated.

An {\it open} system interacts with its surrounding, keeps information on the
preparation of its initial state and, generally, can feel past interactions
through retardation memory effects. Even if, at the given time, the
interactions with its surrounding can be neglected, the system remains open,
if it possesses the memory of its preparation and of past interactions.
In addition, there can exist quantum statistical correlations making the
system entangled with its surrounding.

A physical system is {\it quasi-isolated} when it is isolated from its
surrounding interactions, does not have retardation memory effects, and is
not entangled with surrounding through quantum statistical correlations. Such
a system can be conditionally treated as isolated for a short instance of
time. However, to confirm its isolation, one needs to realize measurements,
at least nondestructive, which perturbs the system making it not absolutely
isolated, but quasi-isolated \cite{Yukalov_15,Yukalov_16,Yukalov_17}.

Contrary to physical systems, a decision maker, even being isolated from
a surrounding society, always keeps memory and information of many previous
interactions. Therefore a decision maker has always to be considered as an
open system.

\section{Composite separable events}

When one deals with not just a single event, but with several events, the
situation becomes more involved. This especially concerns the measurements
of noncommuting observables, whose operators do not commute with each other
\cite{Gudder_18}. There has been an old problem of correctly defining the
quantum probability of such events. It has been shown \cite{Yukalov_19} that
the L\"{u}ders probability \cite{Luders_20} of consecutive measurements is
a transition probability between two quantum states and cannot be accepted
as a quantum extension of the classical conditional probability. Also, the
Wigner distribution \cite{Wigner_21} is just a weighted L\"{u}ders probability
and it cannot be treated as a quantum extension of the classical joint
probability. The correct and most general definition of a quantum joint
probability can be done \cite{Yukalov_19} by employing the Choi-Jamiolkowski
isomorphism \cite{Choi_22,Jamiolkowski_23} expressing the channel-state
duality.

Composite events, that are composed of two or more events, can be separable
or entangled. First, we consider separable events.

Let us be interested in the measurements involving two observables associated
with two operators, $\hat{A}$ and $\hat{B}$, with the related eigenproblems
\be
\label{11}
 \hat A | n \rgl = A_n | n \rgl \; , \qquad
 \hat B | \al \rgl = B_\al | \al \rgl \; .
\ee
As above, we shall denote the events of measuring the eigenvalues $A_n$ and
$B_\alpha$ by the same letters, respectively. To each event, we put into
correspondence the appropriate projectors,
\be
\label{12}
 A_n \ra \hat P_n \equiv | n \rgl  \lgl n| \; , \qquad
B_\al \ra \hat P_\al \equiv | \al \rgl  \lgl \al | \; .
\ee
Constructing two Hilbert space copies
\be
\label{13}
{\cal H}_A \equiv {\rm span} \{ | n \rgl \} \; , \qquad
{\cal H}_B \equiv {\rm span} \{ | \al \rgl \} \;   ,
\ee
we define the algebras of observables
\be
\label{14}
 {\cal A} \equiv \{ \hat P_n \} \; , \qquad
{\cal B} \equiv \{ \hat P_\al \} \; ,
\ee
acting on the corresponding Hilbert spaces. The composite algebra
$\mathcal{A} \bigotimes \mathcal{B}$ acts on the tensor-product space
$\mathcal{H}_A \bigotimes \mathcal{H}_B$. The system statistical state
(decision-maker strategic state) also acts on the space
$\mathcal{H}_A \bigotimes \mathcal{H}_B$. The joint probability of events,
corresponding to the observables from the algebra
$\mathcal{A} \bigotimes \mathcal{B}$, is defined as
\be
\label{15}
 p \left ( {\cal A} \bigotimes {\cal B} \right ) \equiv
{\rm Tr}_{AB} \; \hat\rho {\cal A} \bigotimes {\cal B} \;  ,
\ee
with the trace over $\mathcal{H}_A \bigotimes \mathcal{H}_B$.

For any two operators from the algebra $\mathcal{A}$, it is possible to
introduce the Hilbert-Schmidt scalar product that is a map
\be
\label{16}
 \sgm_A : \; {\cal A} \times {\cal A} \ra \mathbb{C} \; .
\ee

Thus, for the operators $\hat{A}_1$ and $\hat{A}_2$ from the algebra
$\mathcal{A}$, the scalar product reads as
\be
\label{17}
 (\hat A_1 , \hat A_2) \equiv  {\rm Tr}_{A} \hat A_1^+ \hat A_2 \; ,
\ee
with the trace over $\mathcal{H}_A$. The operators from the algebra
$\mathcal{A}$, acting on the Hilbert space $\mathcal{H}_A$, and complemented
with the scalar product $\sigma_A$, form the Hilbert-Schmidt operator space
\be
\label{18}
 \widetilde {\cal A} \equiv \{ {\cal A} , {\cal H}_A , \sgm_A \} \;  .
\ee
Similarly, one defines the Hilbert-Schmidt space
\be
\label{19}
   \widetilde {\cal B} \equiv \{ {\cal B} , {\cal H}_B , \sgm_B \} \; .
\ee
The tensor-product of the above Hilbert-Schmidt spaces forms the space
\be
\label{20}
\widetilde{\cal{C}} \equiv
\widetilde{\cal{A}} \bigotimes \widetilde{\cal{B}} \; .
\ee

An operator $\hat{C}$ acting on $\widetilde{\mathcal{C}}$ is called {\it separable}
if and only if it can be represented as a sum
\be
\label{21}
 \hat C =  \sum_i \hat A_i \bigotimes \hat B_i \qquad
(\hat A_i \in \widetilde{\cal A} , \; \hat B_i \in \widetilde{\cal B} ) \;  .
\ee

A composite event is named a prospect. The prospect $A_n \bigotimes B_\alpha$
corresponds to the prospect operator
\be
\label{22}
\hat P\left ( A_n \bigotimes B_\al \right ) =
\hat P_n \bigotimes \hat P_\al \;   .
\ee
The prospect operator (22) is evidently separable. Therefore the corresponding
prospect $A_n \bigotimes B_\alpha$ is also called separable. The related
prospect probability writes as
\be
\label{23}
 p \left ( A_n \bigotimes B_\al \right ) =
{\rm Tr}_{AB} \; \hat\rho \hat P_n \bigotimes \hat P_\al \;  ,
\ee
with the trace over $\mathcal{H}_A \bigotimes \mathcal{H}_B$.

More generally, the prospect $A_n \bigotimes \bigcup_\alpha B_\alpha$, where
$\bigcup_\alpha B_\alpha$ is a union of mutually orthogonal events, corresponds
to the prospect operator
\be
\label{24}
  \hat P \left ( A_n \bigotimes \bigcup_\al B_\al \right ) =
\sum_\al  \hat P_n \bigotimes \hat P_\al \; .
\ee
This operator is separable, and the related prospect probability
\be
\label{25}
p \left ( A_n \bigotimes \bigcup_\al B_\al \right ) =
\sum_\al  p\left ( A_n \bigotimes B_\al \right )
\ee
is additive with respect to the events $A_n \bigotimes B_\alpha$.

\section{Composite entangled events}

An operator $\hat{C}$ from the Hilbert-Schmidt space (20) is termed
{\it entangled}, or non-separable, if it cannot be represented as sum (21),
so that
\be
\label{26}
  \hat C \neq \sum_i \hat A_i \bigotimes \hat B_i \qquad
( \hat A_i \in \widetilde{\cal A} , \; \hat B_i \in \widetilde{\cal B} ) \;  .
\ee

The appearance of entangled events, corresponding to entangled operators,
is connected with the existence of not operationally testable measurements
that also are called uncertain measurements, or incomplete measurements, or
indecisive measurements, or inconclusive measurements. Respectively, one can
keep in mind uncertain events in decision making.

Let us define an uncertain event $B$ as a set of possible events $B_\alpha$,
characterized by weights $|b_\alpha|^2$,
\be
\label{27}
 B = \{ B_\al : \; \al = 1,2, \ldots \} \;  .
\ee
Since the uncertain event is not operationally testable, it is not required
that the weights $|b_\alpha|^2$ be summed to one. The uncertain-event
operator is
\be
\label{28}
  \hat P(B) = | B \rgl \lgl B | \; ,
\ee
with the state
\be
\label{29}
  | B \rgl = \sum_\al b_\al | \al \rgl
\ee
that is not necessarily normalized to one. The uncertain-event operator (28),
which can be written as
\be
\label{30}
 \hat P(B) = \sum_{\al\bt} b_\al b_\bt^* | \al \rgl \lgl \bt | \;  ,
\ee
is not a projector onto a subspace that would correspond to a degenerate
spectrum, because it does not have form (5). Moreover, it is not a projector
at all, as far as it is not necessarily idempotent,
$$
  \hat P^2(B) = \lgl B | B \rgl \hat P(B) \neq \hat P(B) \; ,
$$
since state (29) is not, generally, normalized to one.

A composite event, formed of an operationally testable event $A_n$ and an
uncertain event (27), is the uncertain prospect
\be
\label{31}
  \pi_n = A_n \bigotimes B \; ,
\ee
whose prospect operator is
\be
\label{32}
  \hat P(\pi_n) \equiv | \pi_n \rgl \lgl \pi_n | =
\sum_\al | b_\al|^2 \hat P_n \bigotimes \hat P_\al +
\sum_{\al\neq\bt} b_\al b_\bt^* \hat P_n \bigotimes | \al \rgl \lgl \bt | \; .
\ee
This operator is not separable in the Hilbert-Schmidt space (20), because
the operator $\vert \alpha \rangle \langle \beta \vert$ does not pertain to
space (19) composed of projectors $\hat{P}_\alpha$. Hence, operator (32)
is entangled. The corresponding prospect (31) is termed entangled.

The prospect operators are assumed to satisfy the resolution of unity
\be
\label{33}
 \sum_n \hat P(\pi_n) = \hat 1 \;  ,
\ee
where $\hat{1}$ is the identity operator in the space
$\mathcal{H}_A \bigotimes \mathcal{H}_B$. But these prospect operators are
not necessarily orthogonal, since
$$
 \hat P(\pi_m) \hat P(\pi_n) =
\lgl \pi_m | \pi_n \rgl | \pi_m \rgl \lgl \pi_n | \;  ,
$$
and they are not idempotent, as far as
$$
  \hat P^2(\pi_n) = \lgl \pi_n | \pi_n \rgl \hat P(\pi_n) \;  .
$$
That is, they are not projectors. The family $\{\hat{P}(\pi_n)\}$ of such
positive operators, obeying the resolution of unity (33) is named
{\it positive operator-valued measure} \cite{Benioff_2,Holevo_4,Holevo_5}.

For a given lattice $\{\pi_n\}$ of prospects, the prospect probabilities
\be
\label{34}
 p(\pi_n) = {\rm Tr} \hat\rho \hat P(\pi_n) \;  ,
\ee
where the trace is over $\mathcal{H}_A \bigotimes \mathcal{H}_B$, satisfy
the conditions
\be
\label{35}
\sum_n p(\pi_n) = 1 \; , \qquad 0 \leq p(\pi_n) \leq 1 \;   ,
\ee
making the set $\{p(\pi_n)\}$ a probability measure.

The analysis of the prospect probability (34) results in the following
properties \cite{Yukalov_24,Yukalov_25,Yukalov_26}. The probability can be
written in the form
\be
\label{36}
 p(\pi_n) = f(\pi_n) + q(\pi_n) \;  .
\ee
The first term,
\be
\label{37}
 f(\pi_n) =
\sum_\al | b_\al |^2 p \left ( A_n \bigotimes B_\al \right ) \;  ,
\ee
corresponds to classical probability, possessing the features
\be
\label{38}
 \sum_n f(\pi_n) = 1 \; , \qquad 0 \leq f(\pi_n) \leq 1 \;  .
\ee
The classical term (37) is an objective quantity reflecting the given
properties of the prospect. The second term,
\be
\label{39}
 q(\pi_n) =
\sum_{\al\neq\bt} b_\al b_\bt^* \lgl n \al | \hat\rho | n \bt \rgl \;  ,
\ee
is purely quantum, caused by interference and coherence effects. This quantum
term in the theory of measurements is called interference factor, or coherence
factor, and in decision theory, attraction factor, since it characterizes the
subjective attractiveness of different prospects to the decision maker.

According to the {\it quantum-classical correspondence principle}
\cite{Bohr_27}, when quantum effects disappear, quantum theory should reduce
to classical theory, which implies
\be
\label{40}
 p(\pi_n) \ra f(\pi_n) \; , \qquad q(\pi_n) \ra 0 \;  .
\ee
In general, the reduction of quantum measurements to their classical
counterparts is called decoherence \cite{Zurek_28}.

The quantum factor (39) varies in the range
\be
\label{41}
  -1 \leq q(\pi_n) \leq 1
\ee
and satisfies the {\it alternation law}
\be
\label{42}
  \sum_n q(\pi_n) = 0 \; .
\ee
For a large number of considered prospects $N$, we get the {\it quarter law}
\be
\label{43}
 \lim_{N\ra\infty} \; \frac{1}{N} \sum_{n=1}^N | q(\pi_n) | =
\frac{1}{4} \;  .
\ee

A known example of the arising interference under measurements is provided
by the double-slit experiment \cite{Neumann_1}. The passage of a particle
through one of two slits corresponds to prospect (31). The operationally
testable event $A_n$ is the registration of the considered particle by an
$n$-th detector, while the passage through one of the two slits, either $B_1$
or $B_2$ is described by the uncertain event $B = \{B_1, B_2\}$.

The quantum term is not always nontrivial \cite{Yukalov_29}. To be nonzero,
it requires the validity of two necessary conditions. The first condition is
that the considered prospect be entangled, as described in Sec. 4. The second
necessary requirement is the entanglement of the statistical operator. The
latter is entangled when, e.g.,
\be
\label{44}
 \hat\rho \neq \hat\rho_A \bigotimes \hat\rho_B \; ,
\ee
where
$$
 \hat\rho_A \equiv {\rm Tr}_B \hat\rho \; , \qquad
\hat\rho_B \equiv {\rm Tr}_A \hat\rho \;  ,
$$
that is, when the statistical state of a composite system is not a product
of its partial subsystem states. More generally, the system state is entangled
if it cannot be represented in the form
\be
\label{45}
 \hat\rho \neq
\sum_\al \lbd_\al \hat\rho_{\al A} \bigotimes \hat\rho_{\al B} \;  ,
\ee
where
$$
 \sum_\al \lbd_\al = 1 \; , \qquad 0 \leq \lbd_\al \leq 1 \;  .
$$
Condition (45) is necessary, but not sufficient for the occurrence of a nonzero
term (39).

In quantum theory, entanglement is a well known notion. In decision theory, it
corresponds to the correlations between different possible events that are
perceived by the decision maker.

\section{Conclusion}

We have shown that quantum measurements and quantum decision making can be
described by the same mathematical tools. In both these cases, the problem
of degeneracy can be avoided by employing the von Neumann method of degeneracy
lifting, which is analogous to the Bogolubov quasi-averaging procedure. The
correct definition of quantum probabilities of composite events, called
prospects, is done by using the Choi-Jamiolkowsky isomorphism. This allows us
to describe any type of composite events, including those corresponding to
noncommutative observables.

The notion of uncertain events and measurements makes it feasible to give a
general scheme for describing measurements under uncertainty and decisions
under uncertainty. This is done by means of positive operator-valued measures.
Prospects are classified into two principally different types, separable and
entangled, depending on the structure of the related prospect operators in
the Hilbert-Schmidt space.

The appearance of a quantum interference term in the quantum probability of
composite events is shown to require two necessary conditions, prospect
entanglement and statistical state entanglement.

Classical measurements and classical decision making are particular cases
of the corresponding quantum counterparts. The reduction of quantum
measurements and decision making to the classical limit occurs when the
interference term becomes zero.

The investigation of the analogies between quantum measurements and quantum
decision making suggests the ways of creating artificial quantum intelligence
\cite{Yukalov_11,Yukalov_12} and gives keys for better understanding of
quantum effects in self-organization of complex systems \cite{YS_30}.


\vskip 2cm


\begin{thebibliography}{99}

\bibitem{Neumann_1}
von Neumann J  1955
{\it Mathematical Foundations of Quantum Mechanics}
(Princeton: Princeton University)

\bibitem{Benioff_2}
Benioff P A  1972
{\it J. Math. Phys.} {\bf 13} 908

\bibitem{Holevo_3}
Holevo A S  1973
{\it J. Multivar. Anal.} {\bf 3} 337

\bibitem{Holevo_4}
Holevo A S  2011
{\it Probabilistic and Statistical Aspects of Quantum Theory}
(Berlin: Springer)

\bibitem{Holevo_5}
Holevo A S and Giovanetti V  2012
{\it Rep. Prog. Phys.} {\bf 75} 046001

\bibitem{Yukalov_6}
Yukalov V I and Sornette D  2008
{\it Phys. Lett. A} {\bf 372} 6867

\bibitem{Yukalov_7}
Yukalov V I and Sornette D  2009
{\it Eur. Phys. J. B} {\bf 71} 533

\bibitem{Yukalov_8}
Yukalov V I and Sornette D  2009
{\it Entropy} {\bf 11} 1073

\bibitem{Bohr_9}
Bohr N  1933
{\it Nature} {\bf 131} 457

\bibitem{Bohr_10}
Bohr N  1958
{\it Atomic Physics and Human Knowledge}
New York: Wiley

\bibitem{Yukalov_11}
Yukalov V I and Sornette D  2009
{\it Laser Phys. Lett.} {\bf 6} 833

\bibitem{Yukalov_12}
Yukalov V I and Sornette D  2014
{\it Springer Proc. Phys.} {\bf 150} 37

\bibitem{Bogolubov_13}
Bogolubov N N  1967
{\it Lectures on Quantum Statistics}, Vol. 1.
New York: Gordon and Breach

\bibitem{Bogolubov_14}
Bogolubov N N  1970
{\it Lectures on Quantum Statistics}, Vol. 2.
New York: Gordon and Breach

\bibitem{Yukalov_15}
Yukalov V I  2003
{\it Phys. Lett. A} {\bf 308} 313

\bibitem{Yukalov_16}
Yukalov V I  2012
{\it Phys. Lett. A} {\bf 376}  550

\bibitem{Yukalov_17}
Yukalov V I  2012
{\it Ann. Phys.} {\bf 327} 253

\bibitem{Gudder_18}
Gudder S  1979
{\it Stochastic Methods in Quantum Mechanics}
New York: North-Holland

\bibitem{Yukalov_19}
Yukalov V I and Sornette D  2013
{\it Laser Phys.} {\bf 23} 105502

\bibitem{Luders_20}
L\"{u}ders G  1951
{\it Ann. Phys.} {\bf 15} 663

\bibitem{Wigner_21}
Wigner E  1932
{\it Phys. Rev.} {\bf 40} 749

\bibitem{Choi_22}
Choi M D  1972
{\it Can. J. Math.} {\bf 24} 520

\bibitem{Jamiolkowski_23}
Jamiolkowski A  1972
{\it Rep. Math. Phys.} {\bf 3} 275

\bibitem{Yukalov_24}
Yukalov V I and Sornette D  2010
{\it Adv. Complex Syst.} {\bf 13} 659

\bibitem{Yukalov_25}
Yukalov V I and Sornette D  2011
{\it Theor. Decis.} {\bf 70} 283

\bibitem{Yukalov_26}
Yukalov V I and Sornette D  2014
{\it IEEE Trans. Syst. Man Cybern. Syst.} {\bf 44} 1155

\bibitem{Bohr_27}
Bohr N  1913
{\it Phil. Mag.} {\bf 26} 857

\bibitem{Zurek_28}
Zurek W H  2003
{\it Rev. Mod. Phys.} {\bf 75} 715

\bibitem{Yukalov_29}
Yukalov V I and Sornette D  2014
{\it Top. Cogn. Sci.} {\bf 6} 79

\bibitem{YS_30}
Yukalov V I and Sornette D 2014
{\it Adv. Complex Syst.} {\bf 17} 1450016

\end{thebibliography}
\end{document}